\documentstyle[12pt]{article}
\font\smalli=cmr8 scaled\magstep1
\def\thecaption#1#2{\centerline{\vbox to 1 in{\hsize 5 in \vfill
           {\textindent{#1} \global \advance \baselineskip by -10 pt
          \smalli \noindent#2 }}}\global \advance \baselineskip by 10 pt}

        \newcommand{\be}{\begin{equation}}
        \newcommand{\ee}{\end{equation}}
        \newcommand{\ba}{\begin{eqnarray}}
        \newcommand{\ea}{\end{eqnarray}}
        \newcommand{\ban}{\begin{eqnarray*}}
        \newcommand{\ean}{\end{eqnarray*}}
        \newcommand{\Ba}{\begin{equation} \begin{array}}
        \newcommand{\Ea}{\end{array} \end{equation}}
\def\lesim{\,{\raise-3pt\hbox{$\sim$}}\!\!\!\!\!{\raise2pt\hbox{$<$}}\,}

\def\a{\alpha}
\def\b{\beta}

\def\m{\mu}

\def\L{\Lambda}

\def\ov{\over}

\begin{document}
\baselineskip 21 pt

%%%%%%%%%%%%%%%%%%%%%%%%%%%%%%%%%%%%%%%%%
%
%       Title
%
%%%%%%%%%%%%%%%%%%%%%%%%%%%%%%%%%%%%%%%%%

        \begin{center}
	{\LARGE Neutrino Spin Transitions 
         and the\\ 
          Violation of the
Equivalence Principle.}\\
	\vspace{1.3cm}
	{\large Mou Roy\\}
	\vspace{.4cm}
	{\small {\it Department of Physics \\
         University of California, Riverside \\
        California 92521-0413, U.\ S.\ A.  \\ }}
         \vspace{1.8cm}
\end{center}

\begin{abstract}

The violation of the equivalence principle (VEP) causing neutrino 
oscillations
is of current interest. We study here the possibility of not only 
flavor
oscillation but spin flavor oscillation of 
ultra high energy ($ \sim$  1 PeV) neutrinos emanating from AGN
due to VEP and due to the presence of a large magnetic
field ($ \sim$ 1 Tesla) in AGN. 
In particular we look at the resonance spin flavor conversion driven
by the AGN potential.
Interesting bounds
on the transition magnetic moment of neutrinos  may therefore be 
obtained.

\end{abstract}

\newpage

The equivalence principle can be stated,``In any and every local 
Lorentz
frame, anywhere and anytime in the Universe, all the 
(non-gravitational)
laws of physics must take on their familiar 
special-relativistic forms."
\cite{meisner}. It follows from this principle that if the 
gravitational
couplings of differently flavored neutrinos are different, 
the equivalence
principle is violated.
Neutrinos with non-zero magnetic moment (transition magnetic 
moment when
looking at different flavors) may undergo spin oscillations 
in the 
presence of a magnetic field. 
Both the effects, ``violation of the equivalence principle"
and ``large neutrino magnetic moments" are speculative 
phenomena.
Nonetheless both have considerable importance when studying 
physics beyond the
standard model.

Active Galactic Nuclei (AGN) are by far the strongest 
sources of ultrahigh
energy neutrinos in the Universe \cite{berezinsky}, 
producing fluxes
detectable with present technology \cite{tele}.
Neutrinos are expected to be generated
mainly through the $ \pi \rightarrow \mu \rightarrow e $ 
decay chain
implying that we can expect twice as many muon type neutrinos 
as electron type
neutrinos. A negligible number of tau neutrinos are produced in the
AGN environment. The search for such high energy neutrinos by the
neutrino telescopes under construction 
(e.g. AMANDA, NESTOR, BAIKAL etc.\cite{tele})
necessitates a clear picture of the expected neutrino fluxes 
for these
objects.
AGN have luminosities ranging from $10^{42}$
to $10^{48}$ ergs$/$sec, corresponding to black hole masses 
of the order
of $10^4 $ to $10^{10} M_{\odot}$, on the natural assumption 
that they are
powered by Eddington-limited accretion onto the black hole.
The spherical accretion model
(based on  works by Kazanas, Protheroe and Ellison
\cite{{pk},{ke}}) is used in most of the calculations
of the neutrino production in central regions of 
AGN \cite{{zp},{sb}}.
 According to this scenario, close to the black hole 
the accretion flow becomes
spherical and a shock is formed where the ram pressure 
of the accretion flow
is balanced by the radiation pressure.
The distance from the AGN center to the shock, 
denoted as the shock
radius $ {\cal R}$ ($\simeq$ a few Schwarzschild radii)
contains the central engine of AGN.  The shock radius 
is parametrized
\cite{{pk},{ke}}
by $ {\cal R} = x  r_g$ where $ r_g $ is the Schwarzschild radius
of the black
 hole, and $x$  is estimated to be in the range of 
$10$ to $100$ \cite{zp}

The matter density at the shock $\rho({\cal R})$ can be estimated
from the accretion
rate needed to support black hole luminosity, and from the radius and
accretion velocity at the shock \cite{zp}
\be \rho \left( {\cal R} \right)
 \simeq 1.4 \times 10^{33} x^{-{5 \over 2}} L^{-1}_{\rm AGN} Q^{-1}
\;{\rm gm/cm}^3   \label{dena} \ee
where $Q(x) = 1 - 0.1 x^{0.31}$ is the efficiency for converting
accretion power into accelerated particles at the shock \cite{ke},
and $L_{\rm AGN}$ is the AGN continuum luminosity 
in units of ergs/sec.

Magnetic effects around
AGN have important astrophysical consequences.
An estimate of magnetic fields involving the 
``equipartition" condition
for specific models \cite{begelman} is $ B  \sim 10^4 $ G.
 We will be interested in the vicinity of the
horizon where the pressure scale height is $ \sim r_g $. 
In this region
we will assume that $ B $ remains at the above value.
We will analyse the effects of tiny non-universality in 
gravitational couplings of neutrinos on such high energy 
neutrinos.
Minakata and Smirnov \cite{smirnov} have looked into this matter but
here we 
also incorporate the effects of the magnetic field in the 
AGN environment which causes a spin flip in addition to the 
flavor transition due to the violation of the 
equivalence principle (VEP).
It has been found \cite{smirnov} that 
the accuracy of testing the equivalence principle, $ \Delta f $
is improved by 25 orders of magnitude for massless neutrinos
and by 11 orders of magnitude  for massive neutrinos 
due to ultra high
neutrino energies and cosmological distances.

We look for  bounds on $\Delta f $ 
when looking into resonance conditions 
due to VEP leading to 
spin flavor flip. This mechanism of neutrino oscillation  if valid
in AGN environment will add to the testimony that tau neutrinos 
can be detected at energies 1 PeV and beyond \cite{pakvasa} by the
neutrino telescopes under construction and will also 
cause a depletion
of neutrino flux by oscillation to unobservable right 
handed neutrinos
(provided of course the neutrinos have non-zero 
transition magnetic
moment).
A bound on the
transition magnetic moment of the neutrino can be calculated
which allows resonant spin flavor transitions in the AGN 
environment.

It has been discussed at length that non-universality 
 in the gravitational couplings of
 different flavors of neutrinos
can give rise to neutrino flavor 
oscillations \cite{gasperini}.
The solution to the solar neutrino problem by
 VEP has been
studied in some detail \cite{bahcall}. The atmospheric
 neutrino
data has also been looked into using this scenario 
\cite{halprin}.

Here we have given a qualitative study of
 the resonance spin flavor conversion of
ultra high energy neutrinos due to VEP and due 
to the presence of a
large magnetic field in AGN.
  
We will consider the gravitational effects in 
the presence of neutrino
matter and mixing incorporating  the effects of magnetic field
assuming neutrinos to have nonzero transition magnetic moment.
The main focus of our work is the effect of VEP on spin flavor
precession. We find a region of
resonance where this conversion will occur unimpeded.
The physics of this resonance is very similar to 
the MSW resonance \cite{msw}.
Such transitions due to matter induced resonance 
has been studied in
sufficient  detail in reference \cite{lim}. Matter, 
magnetic and strong
gravitational field effects are found to directly 
cause resonant spin flip
using standard neutrino couplings to gravity in 
reference \cite{mou}.
Minakata {\it et.al} \cite{smirnov} give an elaborate 
study of the 
effects on the cosmic high energy neutrinos 
due to the breakdown
of the equivalence principle but do not study 
spin precession.

To illustrate resonant spin flavor precession phenomenon, we examine
for simplicity only two neutrino flavors, i.e. the
$ \nu_e - \nu_\mu $ system. Similar results are 
obtainable for $ \nu_\mu- \nu_\tau $ system.
The non universality of gravitational couplings to the two flavors is
given by $ \Delta f$.
Using the chiral bases 
$\nu_{e_L}, \nu_{\mu_L}, \nu_{e_R}, \nu_{\mu_R} $,
the evolution equation for propagation through 
the AGN environment is given by
\be
 {d \ov dr} \left( \begin{array}{c}
 \nu_{e_L}\\ \nu_{\mu_L}\\  \nu_{e_R}\\  \nu_{\mu_R}
\end{array}
\right ) = H_{\rm eff} \left( \begin{array}{c}
\nu_{e_L}\\ \nu_{\mu_L}\\  \nu_{e_R}\\  \nu_{\mu_R}
\end{array}\right )
\ee
where $ H_{\rm eff}$ is a $4 \times 4$ matrix.

Since we are specifically interested in spin flavor 
precession let us look 
as an example
at the $ 2 \times 2 $ matrix associated with the transition 
$ ( \nu_{eL} \rightarrow \nu_{\mu R} ) $ which is given by

\vspace{0.5cm}
\be H_{\rm eff}=
\left( \begin{array}{cccc}
      \vspace{0.5cm}
{1\over 2 E} \Delta m^2_{12} \sin^2\theta_M + 
V_G \cos^2\theta_G  & \m_{e\mu} B^\ast\\
\vspace{0.5cm}
\m_{e\mu} B & {1\over 2 E} \Delta m^2_{12}+ V_G \sin^2 \theta_G
\end{array} \right )
\vspace{0.5cm}
\label{gau}
\ee
where $\theta_M $ is the neutrino mixing angle, 
$\theta_G $ is the gravitational mixing angle and 
$ \Delta m_{12}^2 = m_1^2 - m_2^2$.  
$ B$ is the magnetic field in AGN assumed to be constant
over a scale length of the order $ r_g $ and $ \mu_{e \mu} $ is the
corresponding transition magnetic moment. The second 
term in the diagonal elements is due to the VEP, 
where $ V_G = { 1 \ov 2 } \Delta f E \phi(r) $
\cite{{gasperini},{smirnov}}, $ \phi(r) $
being the gravitational potential and $E$ the neutrino energy.
An order of magnitude estimate below shows that the matter
effects are negligible compared to this term and so we have
not taken them  into consideration.

To compare the orders of magnitude of the matter 
effects and gravitational
effects, we consider the spherical accretion model of AGN , 
density $ \rho$ for typical
cases is found to be $ 10 - 10^4 {\rm eV}^4 $ (\ref{dena}). 
The matter effect is given by a term
of the form
\be
{ G_F \rho \ov m_p} \sim 10^{-33} \rho \;{\rm eV} 
\sim 10^{-29}-10^{-32} \;{\rm eV}
\label{matter}
\ee
where $ G_F$ is the Fermi coupling constant, $ m_p $ 
the proton mass and we
have used typical values of AGN matter density as given
 above. To look at the
gravitational part we need to know values of 
gravitational potential
$ \phi (r) $ and also $ \Delta m_{12}^2 $. 
For an order of magnitude
estimation, let us look at the AGN potential 
at the neutrino production
point.
$ \phi(r) =  M G / r $ is the gravitational potential 
at a radius r from
an object of mass M in the Keplerian approximation. 
At the site of neutrino
production i.e. at $ r = (10-100) r_g $, $ \phi(r) 
\sim 5 \times 10^{-3} $
\cite{smirnov},
for a $ 10^{8} M_{\odot} $ black hole.
So for neutrino energy $ E = 10^{15} {\rm eV} $ (1 PeV),
 the gravitational part gives
\be 
E \Delta f \phi(r) \sim 10^{12} \Delta f \;{\rm eV}
\label{gravity}
\ee
The optimal sensitivity on VEP one would achieve 
by solar neutrino observations
by using next generation water cherenkov detectors 
is $|\Delta f | \sim 10^{-15}- 10^{-16} $ \cite{minakata}.  
This gives the gravitational part to be 
$ \sim 10^{-3} - 10^{-4}\;
{\rm eV} $. Also we will show  that at resonance 
conditions when we are dealing with very high 
energy neutrinos (1 PeV), 
$ \Delta f \sim 10^{-28} \Delta m_{12}^2 / {\rm eV}^2 $ 
which could give
$ \Delta f \sim 10^{-38} $ for extremely 
small value of $ \Delta m_{12}^2 = 10^{-10} {\rm eV}^2 $ 
(vacuum neutrino oscillation scenario).
In that case an estimate of the gravitational 
part would give  $ \sim 10^{-26} {\rm eV} $. 
Clearly in the above two cases the VEP 
effect dominates over the matter 
effects and so normal  matter induced 
resonance effects are suppressed.
To study the dominance of the VEP part at 
this high energy
we have looked at the highest sensitivities predicted till
date for $ \Delta f $, the conclusion clearly holds for
lower sensitivities that is larger values of $ \Delta f $.
We can therefore neglect the matter effects 
on neutrinos in this scenario.

We are looking at non-universality of gravity 
coupling to relevant
flavors and are studying spin transitions. 
Hence  it would be sufficient to
look at the transitions in  (\ref{gau}) for a general idea. 
Looking at the resonance condition by equating the diagonal
terms in  (\ref{gau}) we get,
\be
\phi_{res} \simeq { \Delta m_{12}^2 \ov 2 E^2 \Delta f}
\label{res}
\ee
We have  assumed  that for a fixed 
$ \Delta m_{12}^2 $ and $ V_G $,
$ \cos \theta_m \simeq \cos\theta_G \simeq 1 $.
\footnote {For a more realistic study we should consider different 
values of the mixing angles which has been done in sufficient 
detail in \cite{smirnov} .}
Assuming that the resonance gravitational potential 
is the potential at
the neutrino production site, \footnote
{ At the neutrino production point the AGN potential dominates
over the supercluster and galactic effects \cite{smirnov}. } 
estimated to be $\sim 10^{-3} $, at neutrino
energy being 1 PeV,
\be
\Delta f \sim 10^{-28} \left( { \Delta m_{12}^2 \ov 1 {\rm eV}} \right)
\label{vep}
\ee
This accuracy increases by using still higher energy neutrinos.

The transition matrix can be cast  into  the form
\be \left( \begin{array}{cc}
d & b \\
b & -d
\end{array} \right )    \label{mat}     \ee
the explicit forms of $b$ and $d$ being,
\be
b =  \mu_{e\mu} B \qquad
d = \left(\Delta m^2 {\rm\ term} \right)
 + \left( VEP {\rm\ term} \right)
\ee

Resonances occur when $d$ vanishes, 
in which case the adiabatic transition probability
is well described (for sufficiently slowly varying $d$ and $b$) by the
Landau-Zener approximation \cite{lz}
\be
P_{LZ} = \exp \left\{ -2 \pi^2 \, {\b^2 \ov \a } \right\}
\ee
where
\be
\beta =  
b \;\vert_{\rm res}\;\;\;\; \a = \dot{d}\;\vert_{\rm res} \label{alph}
\ee
The condition for these resonances 
to induce an appreciable transition
probability (adiabaticity condition) is
\be   \b^2 \ge {\a\over 2\pi^2}. \label{cond} \ee
Also as previously mentioned, in the absence of a 
detailed model we will
assume that the magnetic field remains 
approximately constant inside a 
region of size equal to the pressure scale height 
($ \sim r_g $ in this
case), the average fields of two of these regions 
will be uncorrelated.
Let us define a scale height  $\L$ \cite{mou},
\be
\L=\left| {dr\ov {d(\ln \Delta f  E \phi(r))}} \right|
\ee
over which the magnetic field is assumed to have a constant value 
(estimated to be $ \sim 10^4 {\rm G }$).
At resonance $ \alpha \sim \Delta m^2 / E \Lambda $ and the condition
(\ref{cond}) reduces to
\be
\mu_{e\mu}\ge 
{ 1 \over B }\left | { \Delta m_{12}^2 \over 2 \pi^2 E \Lambda } 
\right|^{1/2} = \mu^{\rm res}_{\rm min} \label{cond2}
\ee
Resonant spin flavor transitions will occur 
therefore provided the transition
magnetic moment satisfies the above bound. 
This constraint is however
independent of the gravitational potential
 and $ \Delta f$. Note that we
have assumed the Keplerian approximation 
where $ d V_G / d r = - V_G / r $.
Also we take $ V_G / r = V_G (r_g)/ r_g $ 
and $ r \sim r_g $ 
since we restrict ourselves to the site of 
neutrino production 
($ r \sim r_g $) within which the magnetic  field is constant.
Taking energy to be 1 PeV and $ B = 10^4 $ G , 
considering a $ 10^{8} M_{\odot}$black hole, 
$  \mu^{\rm res}_{\rm min} 
\simeq 10^{-12} |{\Delta m_{12}^2}|^{ 1 \ov 2}  \mu_B $
(where $\mu_B$ denotes the Bohr Magneton)
which  as an example for $ \Delta m_{12}^2 = 10^{-6} $ 
(for solar large
angle solution) gives, $  \mu^{\rm res}_{\rm min} 
= 10^{-15} \mu_B.$
This lies comfortably inside the direct experimental bounds 
($ \mu_\nu \le 10^{-10} \mu_B $)
as well as the astrophysical limits 
($ \mu_\nu \le 10^{-11}\mu_B$).
In view of this such resonances will induce 
significant transition
probabilities whenever the resonance 
conditions are satisfied.

Thus the violation of the equivalence principle 
when applied to ultrahigh
energy AGN neutrinos can effectively induce 
a spin flavor conversion
provided the transition magnetic moment is 
sufficiently large.
Matter effects are found to be negligible in 
comparison to the VEP
terms and so we do not look at usual matter 
induced flavor spin transitions.
However the lack of precise modeling of the 
AGN magnetic field and 
lacking a better understanding of the parameters 
of the neutrino system,
it is impossible to determine unambiguosly whether 
such transitions do take
place.

Ultrahigh energies and strong AGN gravitational potential 
allow to improve
the accuracy of testing the equivalence principle 
by several orders of magnitude as noted 
in \cite{smirnov}. For an energy of  
1 PeV for instance, resonance 
conditions lead to  
$ \Delta f \sim 10^{-28} \Delta m_{12}^2 / 1 {\rm eV}^2 $. 
We have limited our calculations to the 
AGN environment only. A more complete
study would involve the inclusion of the 
intergalactic and galactic potentials
as done in \cite{smirnov} to study the 
total VEP effects on the neutrinos
for their entire distance of travel 
($ \sim 100$ Mpc is a sensible distance
to AGN). But our effort was to show that
 spin flavor transitions can also
be looked  at in such a scenario which 
gives a magnetic moment bound
independent of the gravitational potential. 
Our calculated bound for the transition magnetic moment 
for reasonable values of different parameters is
$  \mu^{\rm res}_{\rm min}
\simeq 10^{-12} |{\Delta m_{12}^2}|^{ 1 \ov 2}  \mu_B $.
 It is of relevance to study
possible helicity flips of neutrinos because 
observable left handed neutrinos
could be converted into unobservable 
right handed ones thereby reducing the
expected neutrino flux.

\pagebreak

\end{document}